\documentstyle[preprint,prl,aps,epsf]{revtex}

\begin{document}
\def\eqn#1{Eq.$\,$({#1})}
\draft
\preprint{}
\title{Numerical renormalization group of vortex aggregation in ${\bf 2D}$
decaying turbulence: the role of three-body interactions}

\author{Cl\'ement Sire and Pierre-Henri Chavanis}
\address{Laboratoire de Physique Quantique (UMR C5626 du CNRS),
Universit\'e Paul Sabatier\\
31062 Toulouse Cedex, France.\\
(clement{@}irsamc2.ups-tlse.fr \& chavanis{@}irsamc2.ups-tlse.fr) \\
(Version of \today)}
\maketitle
\begin{abstract}
In this paper, we introduce a numerical renormalization group procedure which  
permits long-time simulations of vortex dynamics and coalescence in a $2D$ 
turbulent decaying fluid. The number of vortices decreases as $N\sim t^{-\xi}$, 
with $\xi\approx 1$ instead of the value $\xi=4/3$ predicted by a na\"{\i}ve 
kinetic theory. For short time, we find an effective exponent $\xi\approx 
0.7$ consistent with previous simulations and experiments. We show that the
mean square displacement of surviving vortices grows as $\langle
x^2\rangle\sim t^{1+\xi/2}$. Introducing effective dynamics for two-body and
three-body  collisions, we justify that only the latter become relevant at
small vortex  area coverage. A kinetic theory consistent with this mechanism
leads to  $\xi=1$. We find that the theoretical relations between kinetic
parameters are all in good agreement with experiments.
\end{abstract}


\newpage
\tighten


\newpage
\section{Introduction}

In recent years, a great deal of work has been devoted to the study of
two-dimensional turbulence. Two-dimensional turbulence is not only relevant to
the study of geophysical and astrophysical flows,  but it is
also far more accessible to modern computers and experiments, since the
measurement and the visualization of the velocity and  vorticity fields 
are much easier than in $D=3$. In addition, two-dimensional turbulence has
deep connections with other fields of physics such as electron plasmas in
magnetic field \cite{plasma} and stellar dynamics \cite{chstellar}.

For the specific problem of two-dimensional decaying turbulence, recent
experimental \cite{tab1,tab2,tab3} and theoretical
\cite{w1,w2,w3,w4,w5,w6,benzi1,benzi2,benzi3,pomeau,csvor} works have
emphasized the importance of coherent vortex dynamics during the fluid decay.
This process essentially consists in three stages: during an initial transient
period, the fluid self-organizes and a network of coherent vortices appears.
Once the coherent vortices have emerged, vortices disappear  through mergings
of like-sign vortices, such that their number $N$ decreases and their average
radius $r$ increases, in a process somewhat reminiscent of a coarsening dynamics
\cite{csvor}. During this process, and in the limit of small viscosity, energy
remains constant. When only one dipole (or very few)  remains, it finally
decays diffusively.

From the theoretical point of view, the ``coarsening'' stage is certainly the
most interesting as, in principle, it can extend on an arbitrary long time
period. In this regime, the main question arising concerns the existence of
universal features including the decay exponent $\xi$ ($N\sim t^{-\xi}$), and
other exponents which describe the time evolution of quantities such as the
average vortex radius, the enstrophy or the kurtosis.  

In this paper, we first describe an effective model for the vortex dynamics
and review the main experimental and numerical results concerning the temporal
evolution of the physical quantities listed above.  In section {\bf II}, we
show that the surviving vortices have a hyperdiffusive motion with an
effective diffusion coefficient $D\sim t^{\xi/2}$, and a flight time
distribution behaving as $P(\tau)\sim \tau^{\xi/2-3}$. We then consider a
``na\"{\i}ve'' kinetic theory for the vortex decay dynamics predicting
$\xi=4/3$.  In section {\bf III}, we introduce a numerical renormalization
group (RG) procedure which permits very long simulation times. Although the
numerical diffusion coefficient is well described by the preceding kinetic
theory, the decay exponent is found to be significantly lower than expected
($\xi\approx 1$ instead of $\xi=4/3$). In section {\bf IV}, we derive an
effective dynamics for 2 and for 3 neighboring vortices subjected to the
effective noise due to far away vortices. Within these simple models, we
relate the average merging time to the decay exponent $\xi$ found in the RG
simulations. Our main conclusion is that the lower than expected value for
$\xi$ could be explained by the fact that two-body collisions are irrelevant
at large time, whereas three-body collisions predominate. In section {\bf V},
we present a simple kinetic theory taking these three-body collisions into
account and yielding $\xi=1$. The importance of three-body collisions in
vortex dynamics was previously pointed out by Novikov \cite{novikov}, in a
different context. Throughout this paper, we compare our results to recent
experiments \cite{tab1,tab2,tab3} and find a very good overall agreement.

\section{Kirchhoff model}

\subsection{Generalities}

As we are mainly concerned with the coherent vortex dynamics and merging 
processes, it is natural to focus on the effective behavior of the sole 
vortices, neglecting the incoherent background.

The route to such an effective model starts with the work of Kirchhoff 
\cite{kirchoff} who obtained the equations of motion of point-like vortices 
in the zero viscosity limit. Vortices follow a Hamiltonian dynamics where 
the vortex center coordinates $x_i$ and $y_i$ are conjugate variables:
\begin{eqnarray}  
\Gamma_i  \frac{dx_i}{dt}&=&\frac{\partial H}{\partial y_i},\\
\Gamma_i \frac{dy_i}{dt}&=&-\frac{\partial H}{\partial x_i},
\end{eqnarray}
where $\Gamma_i$ is the circulation of vortex $i$, and where $H$ denotes the
following Hamiltonian:
\begin{equation}
H=-\sum_{i\ne j}\Gamma_i\Gamma_j\ln(r_{ij}),
\end{equation}
where $r_{ij}$ stands for the distance between vortices $i$ and $j$. These
equations of motion can be more explicitly written:
\begin{eqnarray}
\frac{dx_i}{dt}&=&-\sum_{j\ne i}\Gamma_j\frac{y_{ij}}{r_{ij}^2},\label{kirch1}\\
\frac{dy_i}{dt}&=&\sum_{j\ne i}\Gamma_j\frac{x_{ij}}{r_{ij}^2}. \label{kirch2}
\end{eqnarray}
These equations are strictly valid for point-like vortices and cannot describe
vortex mergings. This should be accounted by hand by defining {\it ad hoc}
merging rules as introduced in \cite{w2,w4,benzi2}. The authors of \cite{w4}
determined a criterion for the merging of two like-sign vortices of radii
$r_1$ and  $r_2$, within the elliptical-moment model \cite{elliptic}. They
found that, for $r_1\leq r_2$, collapse can be observed for an initial vortex
separation  $d<r_c=ar_2+br_1^2/r_2$, where $a$ and $b$ are numerical constant
of order 1 ($a\approx 2.59$, $b\approx 0.61$ \cite{w4}). Using $r_1<r_2$ and
$a>3b$, one  easily obtains 
\begin{equation} 
\frac{a+b}{2}(r_1+r_2)\leq r_c \leq a(r_1+r_2), 
\end{equation} 
which shows that $r_c$ is of the order of the mean radius. In \cite{benzi2}, the
authors in fact used $r_c=a'(r_1+r_2)$ ($a'\approx 1.7$).  Thus, if one
considers a collection of vortices of the same typical size, it  is clear that
choosing the same critical distance for all mergings, equal to  the average
radius $r$ of the population of vortices, cannot drastically affect the model
properties. This was actually verified in \cite{csvor}.  Now that the merging
criterion has been given, the properties of the vortex resulting from the
merging of two like-sign vortices must be specified.  Motivated by
experiments \cite{tab1,tab2,tab3} and numerical simulations
\cite{w4,benzi1,benzi2}, the authors of \cite{benzi1,benzi2} and \cite{w2,w4}
assumed that the average peak vorticity $\omega$ is conserved throughout the
merging process, as well as the energy since the inviscid limit is considered.
As the total energy scales as $E\sim N\omega ^2r^4$ (with a possible $\ln N$
correction), this shows that $Nr^4$ should be conserved, or equivalently that
a vortex of radius $r'=(r_1^4+r_2^4)^{1/4}$ results from the merging  of two
vortices of radii $r_1$ and $r_2$. Note that this conservation law is
consistent with the observed slow enstrophy dissipation \cite{benzi1}.

Numerical simulations of this model starting from a population of vortices 
having the same typical radius results in a narrow radius distribution at 
all subsequent times \cite{w4,csvor}. Moreover, it is observed that
the number of vortices decays as a power law
\begin{equation}
N(t)\sim \frac{N_0}{\left(1+t/t_0\right)^\xi},
\label{fit}
\end{equation}
with $\xi\approx 0.70-0.75$, much smaller than the exponent predicted by
Batchelor theory ($\xi=2$) \cite{batchelor}. The conservation of the total energy and
mean peak vorticity leads to the occurrence of only one independent exponent
for the time evolution of physical quantities \cite{w2,w4}:
\begin{eqnarray}
&N(t)\sim R^{-2}\sim t^{-\xi},\quad r\sim t^{\xi/4},\label{scaling}\\
&Z\sim t^{-\xi/2},\quad K\sim t^{\xi/2},
\end{eqnarray}
where $R$ is the typical distance between vortices, and $Z$ and $K$ are
respectively the enstrophy and the kurtosis.

The exponent $\xi$ and the predicted scaling laws are consistent with
experiments \cite{tab1,tab2,tab3} and direct numerical simulations \cite{w4}
of Navier-Stokes (NS) equation (using a hyperviscous dissipation term).

\subsection{Limitation of numerical simulations}

The Kirchhoff simulations and actual experiments cited above were only carried
out for very short time. In \cite{w4} (as well as in experiments
\cite{tab1,tab2,tab3}), the number of vortices decays by less than a factor 4
at the maximal accessible time $t_{\rm max}$. Raw data show significant
curvature on a log-log plot, hence the introduction of an extra fitting
parameter $t_0$ in \cite{w4} (see \eqn{\ref{fit}}). Since $t_0\approx t_{\rm
max}/3$, the simulation time is of the order of the transient time $t_0$, and
the scaling regime for $t\gg t_0$ is probably not reached. As a  matter of
fact, the actual exponent $\xi$ obtained by measuring the logarithmic slope
at the final time is of order $\xi\approx 0.6$, as obtained  by Benzi and
co-workers \cite{benzi2,benzi3} (see also section {\bf III.B})

It would be interesting to explore the domain of lower vortex density $n$, as 
in all simulations performed so far the mean free path (of the order of the 
typical distance between vortices \cite{w6,chacs}) remains of the same order 
as the radius size. In other words, the fraction of area occupied by the
vortices,
\begin{equation}
s=\frac{N\pi r^2}{L^2}=n\pi r^2=\pi\left(\frac{r}{R}\right)^2,
\end{equation}
remains quite large in the early time of the dynamics. Because of the scaling
laws of \eqn{\ref{scaling}}, $R$ grows faster than the mean radius $r$, such
that vortices become effectively more and more point-like, a regime which
seems to be out of numerical reach, and which should develop for $t\gg t_0$.

\subsection{Kinetic theory for the Kirchhoff model}

\subsubsection{Diffusion coefficient}
In the absence of mergings, the chaotic Kirchhoff dynamics is known to lead to
an effective diffusive motion of the vortices \cite{csvor,w6,chacs}. The
diffusion coefficient can be calculated by computing the fluctuation time
$T(v)$ for a given vortex velocity $v$ and averaging the quantity $v^2T(v)/4$
over the velocity distribution. This calculation has been extensively
described in \cite{chacs} and we present here a simple heuristic
argument leading more directly to the same result.

Using \eqn{\ref{kirch1}} and \eqn{\ref{kirch2}}, the average velocity squared
is
\begin{equation}
\langle v^2\rangle\sim\sum_j\frac{\Gamma_j^2}{r_{ij}^2},
\end{equation}
where we have neglected the contribution of off-diagonal terms obtained when
squaring \eqn{\ref{kirch1}} and \eqn{\ref{kirch2}}. If we assume the vortices
to be uniformly distributed on average, and the circulations to be equal up to
their sign, we then obtain
\begin{equation}
\langle v^2\rangle\sim N\Gamma^2\frac{2}{L^2}\int_r^L\frac{x\,dx}{x^2}
\sim 2n\Gamma^2\ln(L/r)\sim n\Gamma^2\ln N,\label{v2}
\end{equation}
where we have introduced the vortex typical radius $r$ as a natural cut-off.
This expression already obtained in \cite{csvor} has been qualitatively
checked in \cite{w6} and is confirmed by our simulations of section {\bf III.B}.

It is then natural to assume that the mean free path $l$ is of order $R$, the
typical distance between vortices, as proved in \cite{chacs}. We then get the
expressions for the mean free time $\tau$ and the diffusion coefficient:
\begin{equation}
l\sim R,\quad \tau\sim \frac{l}{v}\sim (n\Gamma\sqrt{\ln N})^{-1},\quad D\sim
\frac{l^2}{\tau}\sim \Gamma\sqrt{\ln N},\label{kinpar}
\end{equation}
in agreement with our more sophisticated treatment \cite{chacs}.

Now, if we include merging events, these different quantities are expected to
vary with time as both the density $n$ and the typical circulation $\Gamma$
do. If we drop logarithmic corrections for now, we obtain
\begin{equation}
l\sim t^{\xi/2},\quad \tau\sim t^{\xi/2},\quad v\sim {\rm const.},\quad D\sim
t^{\xi/2}.\label{scalkin}
\end{equation}
Note that this expression for the diffusion coefficient $D$ differs from that
obtained in \cite{tab3} ($D\sim t^{3\xi/4}$). Indeed, the authors of \cite{tab3}
used the merging time $\tau_{\rm merg.}$ to compute $D$, instead of the
fluctuation or mean free time which is relevant here \cite{chacs,w6}. A
na\"{\i}ve estimate of this merging time is addressed in the next subsection.

Finally, we predict that the mean square displacement of surviving vortices
(or test particles) in the decaying fluid should behave as \cite{chacs}
\begin{equation}
\langle x^2\rangle \sim t^\nu,\quad{\rm with}\quad\nu=1+\frac{\xi}{2}.
\label{nu}
\end{equation}
In \cite{tab3}, the authors found $\xi\approx 0.7$, which leads to
$\nu\approx 1.35$, using \eqn{\ref{nu}}. This must be compared with the
experimental value $\nu\approx 1.3$ for vortices and $\nu\approx 1.4$ for test
particles moving along the current lines of the fluid. \eqn{\ref{nu}} is also
in good agreement with our simulations of section {\bf III.B}.

Note that this hyperdiffusive behavior can be interpreted by invoking a
power-law decreasing flight time distribution (time between two deflections or
large velocity fluctuations). Let us assume that this distribution behaves as
\begin{equation}
P(\tau)\sim\tau^{-\mu}, \qquad{\rm for}\quad \tau\to +\infty,
\end{equation}
with $\mu>2$, such that the average fluctuation time exists. Then, after a
time $t=\sum_{i=1}^m\tau_i\sim m\langle \tau\rangle$ ($m$ deflections), and
using that  $\langle v^2\rangle$ is essentially constant, the mean square
displacement reads
\begin{equation}
\langle x^2\rangle \sim\langle v^2\rangle \left\langle \sum_{i=1}^m\tau_i^2
\right\rangle \sim m\int^t\tau^{2-\mu}\,d\tau \sim t^{4-\mu}.
\end{equation}
This shows that within this interpretation $\nu=4-\mu$, or
\begin{equation}
\mu=3-\frac{\xi}{2}.
\end{equation}
Using $\xi\approx 0.7$, we find $\mu\approx 2.65$, in good agreement with the
experimental value $\mu=2.6\pm 0.2$ measured in \cite{tab3}.

\subsubsection{Na\"{\i}ve kinetic theory}

The merging time $\tau_{\rm merg.}$ is the typical time between two merging
events involving the same vortex. If we assume that vortex mergings occur
whenever two like-sign vortices stand at a distance less than $r_c\sim r$, a
classical cross-section argument leads to
\begin{equation}
\tau_{\rm merg.}\sim (nvr)^{-1}.\label{taumerg}
\end{equation}
If we assume a scaling regime where $n(t)$ decays as a power law, $\tau_{\rm
merg.}$ must behave linearly with time \cite{trizac} as
\begin{equation}
\frac{dn}{dt}\sim -\xi\frac{n}{t}\sim -\frac{n}{2\tau_{\rm merg.}}.
\end{equation}
Using the scaling equations \eqn{\ref{scaling}} and \eqn{\ref{scalkin}}, and
the above expression for $\tau_{\rm merg.}$, we finally obtain 
\begin{equation}
\tau_{\rm merg.}\sim v^{-1}t^{\frac{3}{4}\xi}\sim t^{\frac{3}{4}\xi},
\label{taumergsca}
\end{equation}
neglecting logarithmic corrections in $v$. The constraint $\tau_{\rm
merg.}\sim t$ then leads to $\xi=4/3$, well above the measured value
$\xi\approx 0.7$. Note that our argument is fully consistent with a direct
simulation performed by Trizac \cite{trizac}, who  found $\xi\approx 0.7$, in
a balistic system obeying the same conservation  laws as in the vortex model,
but for which the typical velocity decreases as  $v\sim t^{-0.47}$  instead of
being constant. Using \eqn{\ref{taumergsca}}, we  indeed predict a decay
exponent $\xi=\frac{4}{3}(1-0.47)\approx 0.707$, in  perfect agreement with
the observed decay exponent. Note that the similarity with the value of the
decay exponent observed in vortex dynamics is then purely incidental.

In the experiment \cite{tab3}, the authors obtained $\tau_{\rm  merg.}\sim
t^{0.6}$, which strongly indicates that they had not yet reached the scaling
regime, which is not too surprising as the experiment was performed on less
than a time decade. As mentioned in section {\bf II.B}, this should raise
some doubts about the validity of the apparent exponent $\xi\approx 0.7$.
  
\section{Numerical renormalization group (RG)}

\subsection{Implementation of the numerical RG}
In this section, we address the problem of performing long-time simulations
allowing for low vortex density {\it and} total vortex area coverage. Direct
simulations of Kirchhoff equations are doomed to failure since each evaluation
of a vortex velocity involves the sum over $N$ terms (see \eqn{\ref{kirch1}}
and \eqn{\ref{kirch2}}). To access very  low densities, one must therefore
start out with a very large initial number of vortices,  which results in a
very slow early dynamics.

The idea of the new approach that we introduce in this section, is to work
with a {\it constant} number of vortices in a reasonable range ($N\sim
20-100$), and progressively increase the domain size $L$ by a procedure
detailed below. For simplicity, we work with vortices of identical radii at
all times and thus of equal circulation up to a sign. This is probably a
reasonable approximation, as previous short-time simulations and experiments
have shown that the radius and circulation distributions remain narrow at all
times \cite{w4,csvor,trizac}. Therefore, the universal features of the
dynamics are expected to survive. Moreover, keeping such constant
distributions in time should minimize transient time effects due to the fact
that the scaling distributions are not yet reached (although this problem is
partly taken into account by the clever procedure introduced in \cite{w4}).

We thus consider $N$ vortices of radius $r$ and circulation $\Gamma=\pm
\omega\pi r^2$ ($N/2$ vortices of each sign) in a box of initial linear size
$L$, such that the initial density is $n_0=N/L^2$. Periodic boundary
conditions are considered and Kirchhoff equations of motion are adapted
to this situation \cite{w3,w4}. Compared to \eqn{\ref{kirch1}} and
\eqn{\ref{kirch2}}, the velocity induced by a vortex $j$ on a vortex $i$ is
only significantly different for vortices at a distance of order $L$, so that
the physics is not modified particularly for vortices at a typical distance
$L/\sqrt{N}$ or less.

Vortices do obey Kirchhoff dynamics until two like-sign vortices meet, {\it 
i.e.} their distance is less than $r_c=2r$ (see the discussion of section 
{\bf II.A}). Both vortices are merged, and all radii and circulations are 
updated to 
\begin{equation} 
r'=\left(\frac{N}{N-1}\right)^{\frac{1}{4}}r,\quad \Gamma'=\pm \omega\pi 
r'^2. \end{equation} The density is updated accordingly: \begin{equation} 
n'=\frac{N-1}{N}n. 
\end{equation} 
A new vortex of the same sign as the vortex which just disappeared is then 
introduced in the box at a random position and has radius $r'$ and
circulation $\Gamma'$. The number of vortices in the box is then restored to
its  initial value $N$. All the distances are then scaled by a factor $\left(
\frac{N}{N-1} \right)^{1/2}$: 
\begin{equation} 
L'=\left(\frac{N}{N-1}\right)^{\frac{1}{2}}L,\quad {\bf
r}'_i=\left(\frac{N}{N-1}\right)^{\frac{1}{2}}{\bf r}_i.  
\end{equation} 
This renormalization procedure ensures that the new density takes the correct 
value 
\begin{equation} 
n'=\frac{N}{L'^2}=\frac{N-1}{N}\frac{N}{L^2} =\frac{N-1}{N}n, 
\end{equation} 
and that the quantity $n'r'^4=nr^4$ is conserved, ensuring the conservation 
of the energy per area unit. If $N$ is large enough, one may expect that the 
introduction of a new uncorrelated vortex after each merging should not 
affect the dynamics, especially at large times for which the merging time is 
much larger than the mean free time. In this regime, the newly introduced 
vortex which has only a probability $N^{-1}$ of being involved in the next 
collision has plenty of time to get ``randomized'' as $\tau_{\rm merg.} 
\gg\tau$.

\subsection{Numerical results}
We have performed long-time RG simulations with $N=10,20,40,60,80$ vortices, 
reaching final densities as low as $2\times 10^{-4}n_0$. Except for the case 
$N=10$ which seems to decay faster, the different density plots are 
essentially independent from the actual number of vortices involved in the RG 
(see fig. 1). The long-time decay exponent is estimated to be $\xi=0.99 \pm 
0.01$ significantly higher than the expected value $\xi\approx 0.7$ but still 
well below the na\"{\i}ve estimate $\xi=4/3$ obtained in section {\bf II.C}.

We have also measured the average mean square displacement of surviving
vortices. The motion is found to be hyperdiffusive with a diffusion exponent
$\nu$ consistent with the prediction of section {\bf II.C} and \cite{chacs}.
Indeed, we find $\nu=1.50\pm 0.01$, to be compared with $\nu=3/2$, admitting
the value $\xi=1$. This is illustrated in fig. 2. Note that according to
section {\bf II.C.}{\it 2}, we thus predict a  flight time distribution
behaving as $P(\tau)\sim\tau^{-5/2}$, for large $\tau$.

In the range $n/n_0=0.8-0.2$, corresponding to the range of density obtained 
in previous experiments \cite{tab1,tab2,tab3} and simulations 
\cite{w1,w4,benzi2,csvor}, we indeed obtain an apparent exponent of order 
$\xi\approx 0.7$. In fig. 3, we compare the direct simulations of 
\cite{w4,csvor} where vortices were allowed to develop a scaling radius 
distribution. After fitting an arbitrary time scale, it appears that our RG
simulations are in good agreement with these previous works, although our RG
simulations extend to almost three more decades in time.

We have also measured the mean square vortex velocity which is
expected to behave as
\begin{equation}
\langle v^2\rangle\sim n\Gamma^2\ln\left(\frac{L(t)}{r(t)}\right)
\sim \ln(t).\label{v2log}
\end{equation}
This behavior is confirmed by our RG numerical simulations as shown in 
fig. 4.

We have also performed RG simulations introducing a distance cut-off in
Kirchhoff equations \eqn{\ref{kirch1}} and \eqn{\ref{kirch2}}, replacing
$r_{ij}^2$ by $(r_{ij}^2+r^2)$. Indeed, although like-sign vortices cannot
approach each other closer than a distance $2r$ (otherwise they merge),
opposite-sign vortices can, which is quite unphysical as it generates very
fast traveling dipoles. Introducing this cut-off results in a physical upper
cut-off of order $\Gamma/r$ for the maximum velocity of these dipoles. The
number of vortices initially decays slightly more slowly, although  the
long-time decay exponent remains fully compatible with $\xi\approx 1$.

It is interesting to compare our results to new direct simulations where the
radii and modulus of circulation are maintained equal for all surviving
vortices. Even starting from $N=2000$, it is hard to reach low densities in a 
reliable way. As exemplified in fig. 5, these direct simulations follow our 
RG calculations before decaying faster beyond a breaking time $t_N$, as the 
density approaches the minimum reachable density $n/n_0=2/N$. We observe that 
the breakdown occurs sooner for samples with a decreasing initial number of 
vortices. $t_N$ is in fact of the order of the minimum time after which some 
samples had reached the minimum possible density (only 2 opposite vortices 
left). Still, fig. 5 lends credence to the claim that the large $N$ limit 
finally reproduces the RG results. Note finally that for these direct
simulations, the actual logarithmic correction in the mean square velocity
behaves differently from that in the RG simulations as the box size $L$
remains constant. In this case, it slowly decreases as
\begin{equation} 
\langle v^2\rangle\sim 
n\Gamma^2\ln\left(\frac{L}{r(t)}\right) \sim \ln\left(\frac{t_{*}}{t}
\right),
\end{equation}
up to a time of order $t_{*}\sim t_N$. Note that this subdominant difference
between RG and direct simulations could be responsible for a slight
discrepancy in the apparent decay exponent observed in both kinds of
simulations.

\section{Merging time in effective two-body and three-body dynamics}

\subsection{General form of the merging time}
In this section, we are concerned with an alternative way of evaluating the
merging time $\tau_{\rm merg.}$ as a function of the physical parameters $n$,
$r$, $\Gamma$... On the basis of simple dimensional analysis, $\tau_{\rm
merg.}$ can be expected to be proportional to the mean free time $\tau$ 
multiplied by an arbitrary function of the only dimensionless parameter 
$nr^2$. Expecting power-law behaviors for these quantities, a natural {\it 
ansatz} is
\begin{equation}
\tau_{\rm merg.}\sim\frac{\tau}{(nr^2)^\alpha}\label{taumgen},
\end{equation}
where $\alpha$ is an exponent to be determined. Dropping logarithmic
corrections in $\tau$ (see \eqn{\ref{kinpar}}), we obtain
\begin{equation}
\tau_{\rm merg.}\sim\left(n\Gamma(nr^2)^\alpha\right)^{-1}\sim
\omega^{-1}(nr^2)^{-(1+\alpha)}.
\end{equation}
Imposing that $\tau_{\rm merg.}$ must be proportional to the actual time (see
section {\bf II.C}), and using the scaling equations \eqn{\ref{scaling}}, we
find the relation between the decay exponent $\xi$ and $\alpha$:
\begin{equation}
\xi=\frac{2}{1+\alpha}.
\end{equation}
Note that the na\"{\i}ve expression of $\tau_{\rm merg.}$ obtained in section
{\bf II.C} can also be written in this form:
\begin{equation}
\tau_{\rm merg.}\sim (nvr)^{-1}\sim \omega^{-1}(nr^2)^{-3/2},
\end{equation}
corresponding to $\alpha=1/2$ and thus $\xi=4/3$.

\subsection{Theory of the effective two-body dynamics}

We now consider the effective dynamics of two {\it nearby} like-sign vortices,
assuming that the $(N-2)$ other vortices are at a distance {\it at least of
order} $R$. The velocity induced by the other vortices on one of these two
vortices can be written as the sum of the velocity induced on the center of
mass (at ${\bf r}_0=({\bf r}_1+{\bf r}_2)/2$) plus a small correction:
\begin{eqnarray}
{v_1}_x&=&-\sum_{j\ne 1,2}\Gamma_j\frac{y_{1j}}{r_{1j}^2},\\
&=&{v_0}_x+\delta x_1\sum_{j\ne 1,2}\Gamma_j
\frac{x_{0j}y_{0j}}{r_{0j}^4}+\delta y_1\sum_{j\ne 1,2}\Gamma_j
\frac{y_{0j}^2-x_{0j}^2}{r_{0j}^4},\\
&=&{v_0}_x+\delta x_1\eta_a+\delta y_1\eta_b, \label{vx}
\end{eqnarray}
where we have neglected other corrections involving higher powers of $\delta 
x_1=x_1-x_0$ and $\delta y_1=y_1-y_0$. After a straightforward calculation, 
we get a similar equation for ${v_1}_y$:
\begin{equation}
{v_1}_y={v_0}_y+\delta x_1\eta_b-\delta y_1\eta_a.\label{vy}
\end{equation}
Note the antisymmetric structure of \eqn{\ref{vx}} and \eqn{\ref{vy}} 
resulting from the Hamiltonian nature of the dynamics.

We will now assume that ${v_0}_x$, ${v_0}_y$, $\eta_a$ and $\eta_b$ can be
considered as random Gaussian variables. Their second moments are 
respectively
\begin{equation}
\langle {v_0^2}_x\rangle=\langle {v_0^2}_x\rangle=\frac{\langle 
v^2\rangle}{2}
\sim n\Gamma^2,
\end{equation}
up to a logarithmic term (see section {\bf II.C} and \cite{csvor,w6,chacs}),
and
\begin{equation}
\langle {\eta_a^2}\rangle=\langle {\eta_b^2}\rangle=N\Gamma^2
\left\langle \frac{x^2y^2}{r^8}\right\rangle\sim 
N\Gamma^2\frac{1}{L^2}\int_R^L
\frac{x^4}{x^8}x\,dx\sim n^2\Gamma^2.
\end{equation}
We have used $R$ as the lower cut-off since the other vortices were
assumed to be at a distance at least of order $R$. Note that $\eta_{a,b}$ have
the  dimension of an inverse time and are simply of order $\tau^{-1}\sim
n\Gamma$.  It is natural to assume that $\eta_{a,b}$ have the same correlation
time as  the velocity, namely the mean free time $\tau$.

Assuming that ${v_0}_{x,y}$ and $\eta_{a,b}$ are Gaussian noises of
correlation time $\tau$ leads us to write a simplified effective Langevin
equation describing the dynamics of these four quantities:
\begin{equation}
\frac{du}{dt}=-\frac{u}{\tau}+\frac{\langle u^2\rangle}{\tau^{1/2}}w_u,
\end{equation} 
where $u={v_0}_{x,y},\eta_{a,b}$, and $w_u$ are independent 
$\delta$-correlated white noises. Such a Langevin equation was recently 
introduced to describe velocity fluctuations in \cite{w6}, with reasonable 
success.

We are now ready to construct an effective two-body system in order to study
the merging time. We consider two like-sign vortices of identical circulation
in a square box of size $R$ with periodic boundary conditions. These vortices
are submitted to their mutual advection and to the effective noise
induced by the other vortices at a distance greater than $R$. If we define 
$x=x1-x2$, $y=y1-y2$ and $d=\sqrt{x^2+y^2}$, we get from Kirchhoff
equations and \eqn{\ref{vx}} and \eqn{\ref{vy}}
\begin{eqnarray}
\frac{dx}{dt}&=&-\Gamma\frac{y}{d^2}+x\eta_a+y\eta_b,\\
\frac{dy}{dt}&=& \Gamma\frac{x}{d^2}+x\eta_b-y\eta_a,
\end{eqnarray}
as the average induced velocity cancels out. After expressing time in units of 
$\tau$ ($t\to t/\tau)$ and distance in units of $R$ ($x\to x/R$, $y\to y/R$), 
and noting that $\Gamma\sim R^2/\tau$, we end up with a dimensionless 
equation of motion as anticipated in the previous subsection:
\begin{eqnarray}
\frac{dx}{dt}&=&-\frac{y}{d^2}+x\eta_\alpha+y\eta_\beta,\label{lang1}\\
\frac{dy}{dt}&=& \frac{x}{d^2}+x\eta_\beta-y\eta_\alpha,\label{lang2}
\end{eqnarray}
where $\eta_{\alpha,\beta}$ are independent Langevin random variables, of 
unit mean square average and correlation time.

Both vortices are initially randomly placed  in the unit box at a mutual
distance $d>0.4 R$ and their relative distance evolves according to
\eqn{\ref{lang1}} and \eqn{\ref{lang2}} until $d$ becomes smaller than the
scaled dimensionless  parameter $d_c=2r/R=2\sqrt{nr^2}$, which defines the
merging time. As anticipated above, the average merging time in units of
$\tau$ can only be a function of this parameter $d_c$, leading to
\eqn{\ref{taumgen}}.
 
\subsection{Absence of strictly two-body collisions} 
 
Numerical simulations of \eqn{\ref{lang1}} and \eqn{\ref{lang2}} lead to the
following surprising result: both vortices remain at a relative distance
greater than a constant $d_{\rm min}$ which slightly depends on the actual
numerical constants in \eqn{\ref{lang1}} and \eqn{\ref{lang2}} (to simplify,
we have assumed the coefficients of ${(x,y)}/{d^2}$, $\langle {\eta_{a,b}^2}
\rangle$, and the correlation time of $\eta_{a,b}$ to be exactly equal to 1).
A typical long-time trajectory is shown in fig. 6 and perfectly illustrates
the absence of collisions when the vortex size is below a certain threshold.
Note that if there were no noise due to the other vortices, both vortices
would strictly remain at the same distance, hence producing a circular
trajectory.

Of course, our result does not prove the absence of collision in the actual
$N$-body system, but strongly suggests that the main assumption according to
which all  other vortices are at a distance greater than $R$ prevents both
test vortices from colliding.

Let us now give a physical interpretation of our result. If both test
vortices were at the same point, their distance would not vary since the
velocity induced by the other vortices would be exactly the same for both
vortices. Thus, when they are close to each other, the effective induced noise
is reduced  linearly with their distance $d$ as shown by \eqn{\ref{lang1}} and
\eqn{\ref{lang2}}. In addition, as vortices get closer to each other ($d\sim
r$), their relative position describes a circle, moving at angular velocity of
order
\begin{equation}
\Omega\sim v_r \times r^{-1}\sim \frac{\Gamma}{r}\times r^{-1}\sim {\omega}.
\end{equation}
Because of this fast rotation, the effective fluctuation time of the noise, as
seen in the moving frame, becomes $\omega^{-1}\ll \tau$ instead of $\tau$.
Hence,  not only the driving noise is reduced due to the proximity of the test
vortices, but it is also averaged out due to their fast rotation. Such a
short effective fluctuation time was in fact introduced in \cite{csvor}.
Another way of interprating the effect of the fast vortex rotation is to
note that due to the large difference between the system natural
frequency $\omega$ and that of the noise perturbation
($\omega'\sim\tau^{-1}$), the adiabatic theorem ensures that the effective
perturbation is reduced by a factor of order $\exp(-C\omega\tau)$, where $C$
is a constant of order unity \cite{landau}.

Conversely, if both vortices start at a distance $d < d_{\rm min}$, this fast
rotating pair remains stable for a very long time, probably infinite. Such
pairs can thus only be destroyed by the direct interaction with a third
vortex, and not solely by the background noise.

The results of this section suggest that strictly two-body interactions are
not sufficient to generate collisions. It is thus natural to study the
equivalent three-body problem, which is the subject of the next section.

\subsection{Merging time of a three-body system}

Using \eqn{\ref{vx}} and \eqn{\ref{vy}}, we can generalize our preceding
approach to the case of three test vortices submitted to the effective  noise
induced by far away vortices. For the first vortex, the effective equation of
motion now reads
\begin{eqnarray}
\frac{dx_1}{dt}&=&-\Gamma_2\frac{y_{12}}{r_{12}^2}
-\Gamma_3\frac{y_{13}}{r_{13}^2} +\delta x_1\eta_a+\delta 
y_1\eta_b+{v_0}_x,\\
\frac{dy_1}{dt}&=&\Gamma_2\frac{x_{12}}{r_{12}^2}
+\Gamma_3\frac{x_{13}}{r_{13}^2} +\delta x_1\eta_b-\delta y_1\eta_a+{v_0}_y,
\end{eqnarray}
with similar equations for the two other test vortices. As we are
studying the merging time of like-sign vortices, we take vortex 1 and 2 to be
of circulation $+\Gamma$, whereas vortex 3 is left unspecified, with
circulation $\pm\Gamma$.

As in the preceding section, these effective equations of motion can be
rescaled by expressing distances in units of $R$ and times in units of $\tau$.
In these new units, a collision between vortices 1 and 2 occurs when the
distance between them is less than the scaled dimensionless parameter
$d_c=2r/R$.

The sign of the third vortex plays a significant role. When it is the same as
that of the other two, a phenomenon reminiscent of that which was found for
two-body collisions occurs: vortices do not collide below a certain radius,
at least during numerically observable times.

On the contrary, when this third vortex is of the opposite sign, we observe a
smooth dependence of the merging time as a function of $d_c$, which is fully
compatible with the functional form (see fig. 7)
\begin{equation}
\tau_{\rm merg.}\sim\frac{\tau}{nr^2},
\end{equation}
that is $\alpha=1$, and thus $\xi=1$ using the results of section {\bf IV.A}.
This result is in agreement with our RG calculation, for which we also found
$\xi\approx 1$. Note that for $r/R$ large, corresponding to the early stage of
the actual dynamics, the apparent value of $\alpha$ is of order
$\alpha\sim 2$, which is compatible with an apparent decay exponent in the
range $\xi\sim 0.6-0.7$ (see the inset of fig. 7).

\section{Physical picture}

The present study strongly suggests that for small surface coverage ($r/R\ll
1$), the relevant collision mechanism involves three vortices, one having an
opposite circulation from the other two. A na\"{\i}ve picture would be that of
a $(+\Gamma,-\Gamma)$ dipole moving at the typical velocity $\Gamma/r$
encountering an isolated vortex of any sign. The importance of these fast
traveling pairs was already suggested in \cite{w3,w6}. It is likely that the
three-body collision processes need not to strictly occur in the simple way
described above, although this does provide an evocative physical picture from
which to construct an effective kinetic theory.

Following the interpretation presented above, the collision rate is
\begin{equation}
\frac{dn}{dt}\sim -n_{\rm dip.}\times nv_{\rm dip.}r,
\end{equation}
where $n_{\rm dip.}$ and $v_{\rm dip.}$ are the dipole density and  typical
velocity, and the last term is the probability per unit time for a dipole to
collide with an isolated vortex. In a simple mean-field approach, dipoles of
typical size $r$ are formed with density
\begin{equation}
n_{\rm dip.}\sim n\times nr^2,
\end{equation}
their velocity being of order $v_{\rm dip.}\sim \Gamma/r$. Finally, we obtain
\begin{equation}
\frac{dn}{dt}\sim -n\times  n^2r^2\Gamma,\label{detail}
\end{equation}
which corresponds to a merging time
\begin{equation}
\tau_{\rm merg.}\sim(n\Gamma\times nr^2)^{-1}\sim\frac{\tau}{nr^2},
\label{taufin}
\end{equation}
which in the language of section {\bf II.C} leads to $\alpha=1$, and then to
$\xi=1$ (up to logarithmic corrections).

This simple interpretation reconciles the paradoxical observation that
although vortices are {\it hyperdiffusive} (see section {\bf II.C.}{\it 1}
and {\bf III.B}) the observed decay exponent $\xi$ is {\it lower} than that
found  for diffusive aggregation \cite{csvor,taka,mstaka} (see also section
{\bf  II.C.}{\it 2}). The dynamics is slowed down by the requirement of three
close vortices for an actual merging to occur, at least for small area
coverage. Note that using \eqn{\ref{taufin}} and energy conservation, we
can also write
\begin{equation}
\tau_{\rm merg.}\sim \frac{\omega L^2}{nE}.
\label{tauexp}
\end{equation}
This relation is particularly well obeyed in the experiment described in 
\cite{tab3}, LHS and RHS terms being respectively 
\begin{equation}
\tau_{\rm merg.}\sim t^{0.57\pm 0.12}, \quad {\rm and}\quad 
\frac{\omega L^2}{nE}\sim \frac{t^{-0.15\pm 0.04}}{t^{-0.70\pm 0.1}}\sim 
\sim t^{0.55\pm 0.14}.
\end{equation}
Thus, although $\tau_{\rm merg.}$ does not behave linearly with time in the
experiment, which is probably due to the fact that the scaling regime was not
reached, all the relations found between physical quantities ($D\sim
n^{-1/2}$, $\nu=1+\xi/2$ and $\mu=3-\xi/2$ in section {\bf II.C}, and the
above relation \eqn{\ref{tauexp}}) are fully consistent with our kinetic
theory.

Note that \eqn{\ref{detail}} was obtained by Pomeau \cite{pomeau} and by one
of us \cite{csvor}, but with the use of highly questionable physical arguments.
In \cite{pomeau}, the kinetic equation was written in the form
\begin{equation}
\tau\frac{dn}{dt}\sim -n\times  nr^2,\label{detailpom}
\end{equation}
by arguing that after a time of order $\tau$ the collision probability is
simply the geometrical overlapping probability $nr^2$, although a
cross-section argument is definitely required (see section {\bf II.C}). This
argument boils down to the assumption that after a time $\tau$, the vortex
positions can be considered to be randomly generated, and collisions happen if
two vortices overlap. In \cite{csvor}, inspired by the theory of diffusive
aggregation \cite{taka,mstaka}, the kinetic equation was written in the form
\begin{equation}
\frac{dn}{dt}\sim -Dn^2,\label{detailcs}
\end{equation}
but with an incorrect expression for the diffusion coefficient $D$ (found
constant in \cite{csvor}). In \cite{csvor}, the fact that $\langle
v^2\rangle$ is essentially constant was correctly used, but the fluctuation
time was taken as $\omega^{-1}$ instead of $\tau$. As we have seen in section
{\bf IV.C}, it happens that just before a collision, the effective fluctuation
time in fact becomes equal to $\omega^{-1}$, which makes the agreement of
\eqn{\ref{detailcs}} with \eqn{\ref{detail}} rather incidental.

\section{Conclusion}

In this paper, we have introduced a numerical RG procedure which permits very
long-time simulations of the vortex Kirchhoff dynamics in a two-dimensional
decaying fluid. Although we recover a short-time regime compatible with a
decay exponent of order $\xi\approx 0.7$ when the vortex surface coverage is
still large and $n/n_0\gtrsim 0.2$, we ultimately find a long-time asymptotic
decay with $\xi\approx 1$. None of these results can be explained by the
simple kinetic theory of section {\bf II.C} based on the occurrence of
two-body collisions which predicts $\xi=4/3$. The failure of this
``na\"{\i}ve'' kinetic theory could be explained by our claim that strictly
two-body collisions are irrelevant for small enough vortex surface coverage
(section {\bf IV. C}). For collision processes involving two like-sign
vortices and a third opposite-sign vortex, we found an average merging time
$\tau_{\rm merg.}\sim \tau/(nr^2)$, fully consistent with a decay exponent
$\xi=1$ (section {\bf IV. C}). A simple kinetic theory based on this collision
mechanism also leads to $\xi=1$ and predicts that $\tau_{\rm merg.}\sim
\frac{\omega L^2}{nE}$ in agreement with the experiment described in
\cite{tab3}.  Our prediction \cite{chacs} that the mean square displacement of
surviving vortices goes as $\langle x^2\rangle\sim t^{1+\xi/2}$ is in good
agreement with our RG simulations and with experiments. Moreover, the exponent
describing the decay of the flight time distribution is related to $\xi$ by
$\mu=3-\xi/2$, in perfect agreement with experiment. 

Our work has so far been limited to the study of the dynamics of a population
of vortices {\it having the same radii} at all times. However, it is important
to address the question of the possible dependence of the decay exponent $\xi$
on the form of the radius distribution and/or the initial conditions
\cite{he}, in order to verify its possible universality. Thus, it is a
motivating challenge to generalize our RG approach to the case of a
polydisperse assembly of vortices. A crucial point would be to correctly
specify the radius of the new vortex reinjected after each merging event. This
and the study of the effective three-body dynamics of different
size/circulation vortices should be the subject of a future study 
\cite{cscha2}.

\acknowledgements 

We are grateful to Jane Basson for useful comments on the manuscript, and to
Sidney Redner and Dima Shepelyansky for fruitful discussions. We also thank
B\'ereng\`ere Dubrulle for her interest in this work.


\begin{figure}[ht]
\narrowtext
\epsfxsize=\hsize
\epsfbox{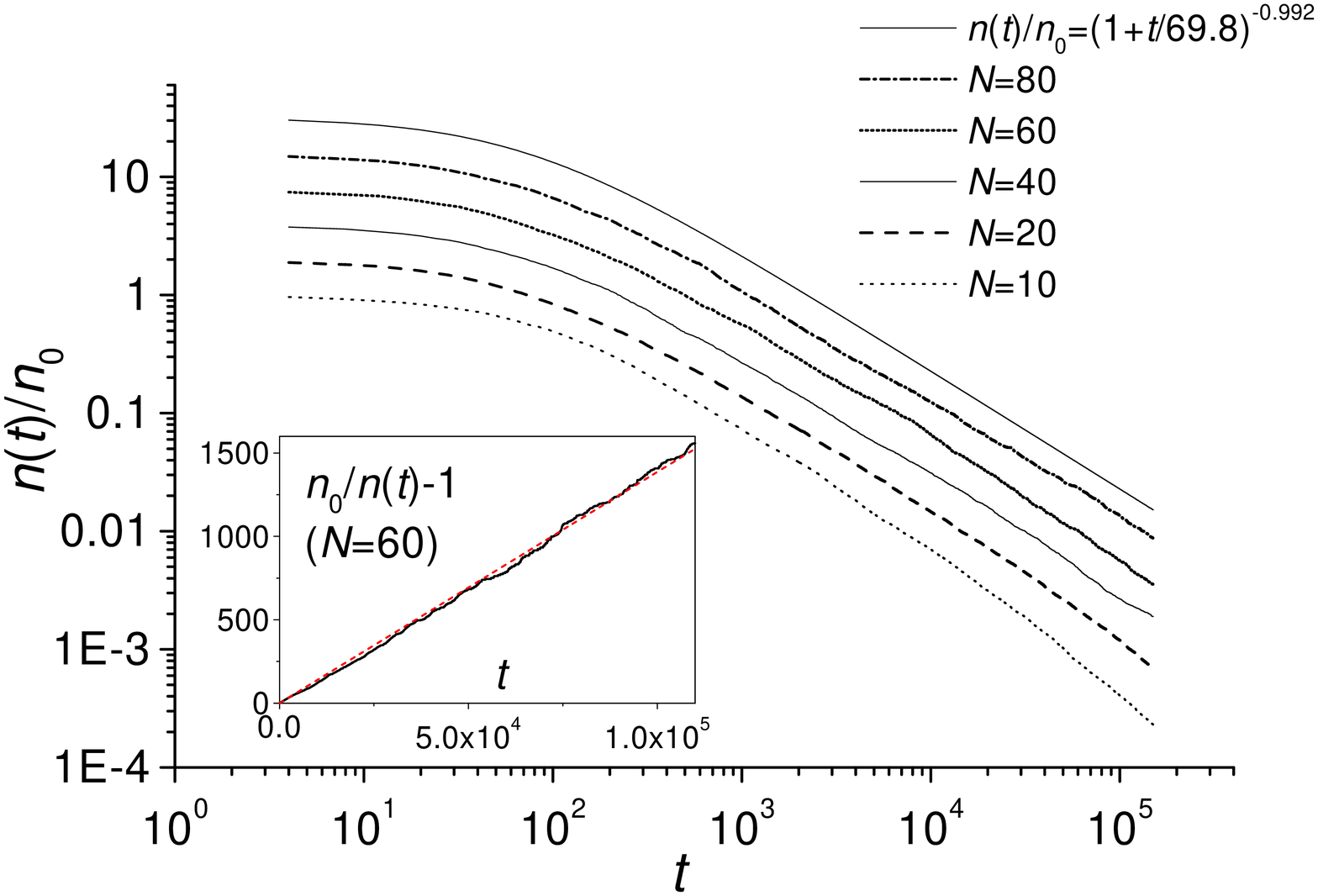} 
\vskip 0cm 
\caption{Numerical RG simulations for $N=10,20,40,60,80$ (resp.
250,150,30,20,10 samples)
and the best fit of the $N=20-80$ curves to the functional form
$n(t)/n_0=(1+t/t_0)^{-\xi}$ (with $\xi=0.992$ and $t_0=69.8$). For clarity,
the curves have been offset by an arbitrary factor 2. The time $t$ is
expressed in units of $\omega^{-1}$ and the initial total area coverage is
10\%, like for most simulations presented in this paper. The inset shows that
$n_0/n(t)-1$ is reasonably linear with time ($N=60$), which is consistent with
an exponent $\xi\approx 1$.}
\end{figure}
\newpage
\begin{figure}[ht]
\narrowtext
\epsfxsize=\hsize
\epsfbox{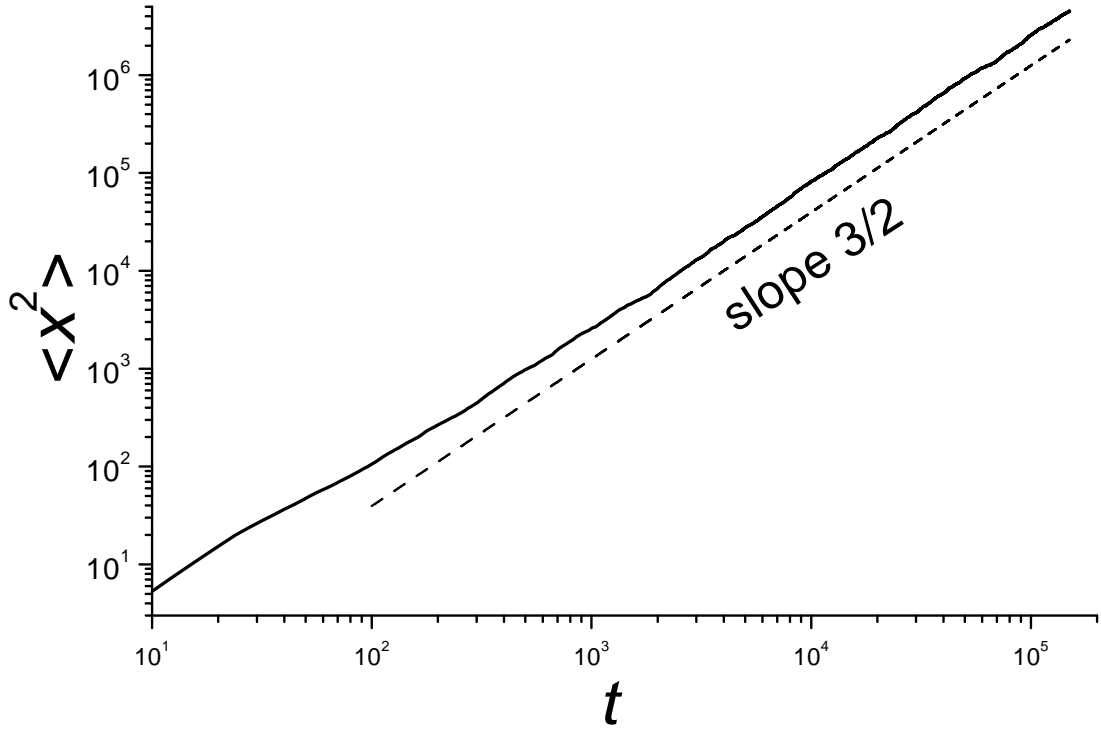}  
\caption{Mean square displacement of surviving vortices as measured in $N=40$
RG simulations. $\langle x^2\rangle\sim t^\nu$, with $\nu=1.50\pm 0.01$, in
perfect agreement with our prediction $\nu=1+\xi/2$ of \eqn{\ref{nu}}, when
taking $\xi=1$.}
\end{figure}
\newpage
\begin{figure}[ht]
\narrowtext
\epsfxsize=\hsize
\epsfbox{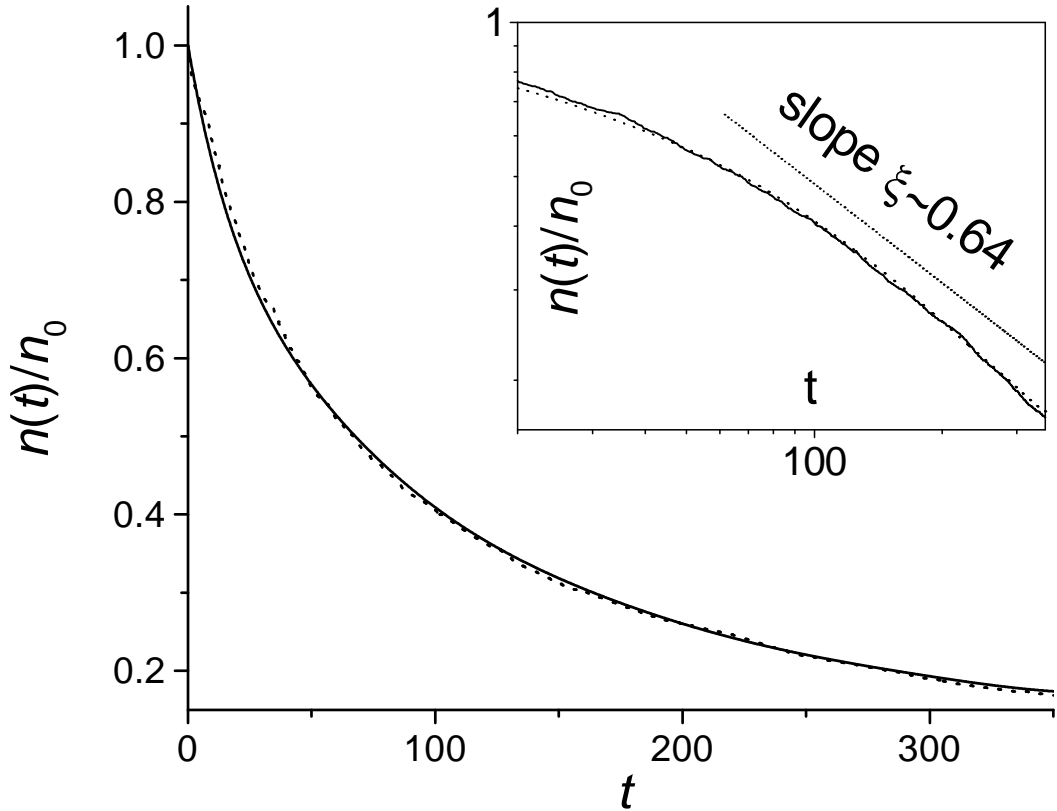} 
\caption{We compare short time RG simulations (dashed curves, $N=60$, 20
samples) with  previous direct simulations including a polydisperse population
of vortices (for which the time unit has been scaled to that
of the RG simulations). The agreement is good and the apparent decay exponent
is of order $\xi\approx 0.64$ (extrapolated to $\xi\sim 0.72-0.75$). However,
the data display a strong curvature in a log-log plot (see inset).}
\end{figure} 
\newpage
\begin{figure}[ht]
\narrowtext
\epsfxsize=\hsize
\epsfbox{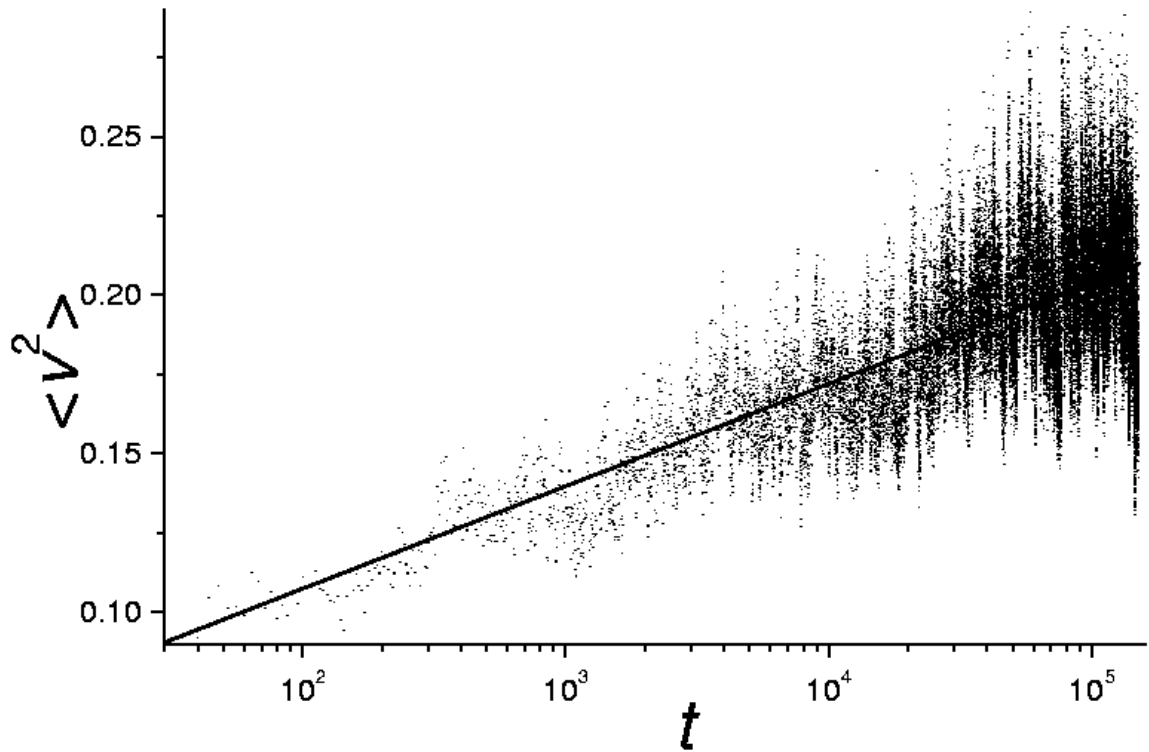}  
\caption{The plot of $\langle v^2\rangle(t)$ (RG simulations with $N=80$; 10
samples; sampling time $\Delta t=4\,\omega^{-1}$, with no time averaging)
displays a slow variation of this quantity fully compatible with the
logarithmic correction obtained in \eqn{\ref{v2log}}. The thick line is
a log-linear fit of the scatter plot.}
\end{figure}
\newpage
\begin{figure}[ht]
\narrowtext
\epsfxsize=\hsize
\epsfbox{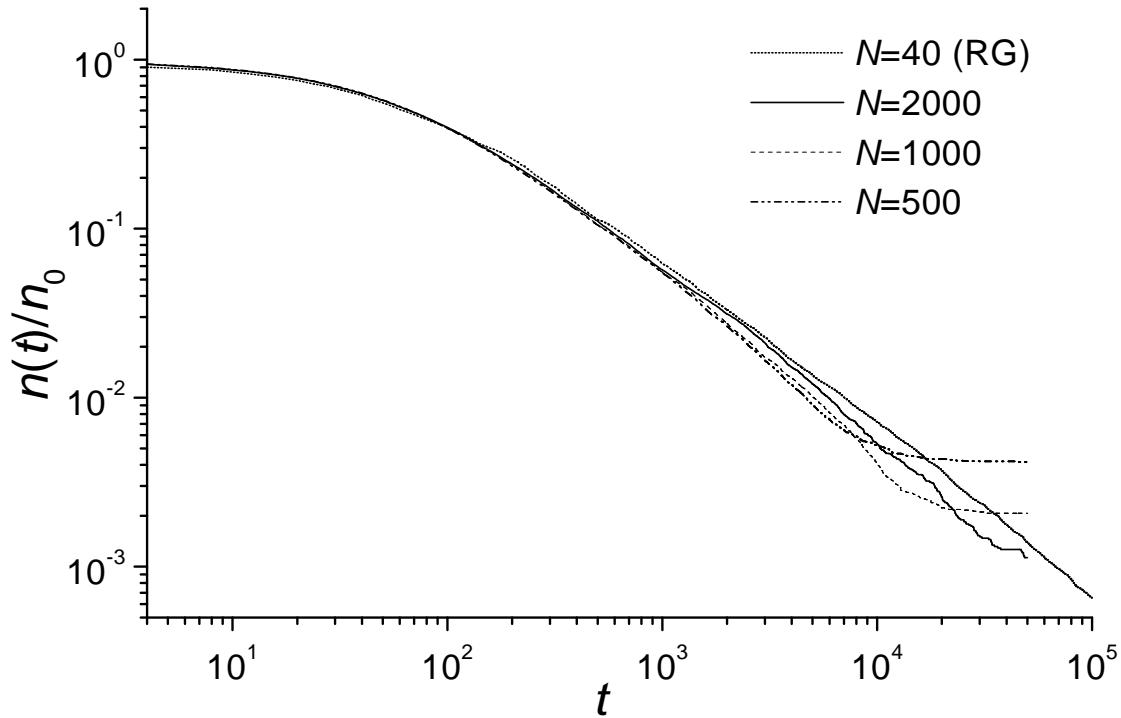}  
\caption{Direct simulations maintaining all radii equal $(N=500,1000,2000)$
are compared to RG simulations $(N=40)$. The saturation to the minimum
reachable density $n/n_0=2/N$ is clearly seen. However, the direct simulations
follow the RG simulations on a longer time domain as $N$ increases, up to a
time $t_N$ for which some samples have already reached the minimum density.
The long time apparent exponent for $N=2000$ is of order $\xi\approx 1.1$.}
\end{figure}
\newpage
\begin{figure}[ht]
\narrowtext
\epsfxsize=\hsize 
\epsfbox{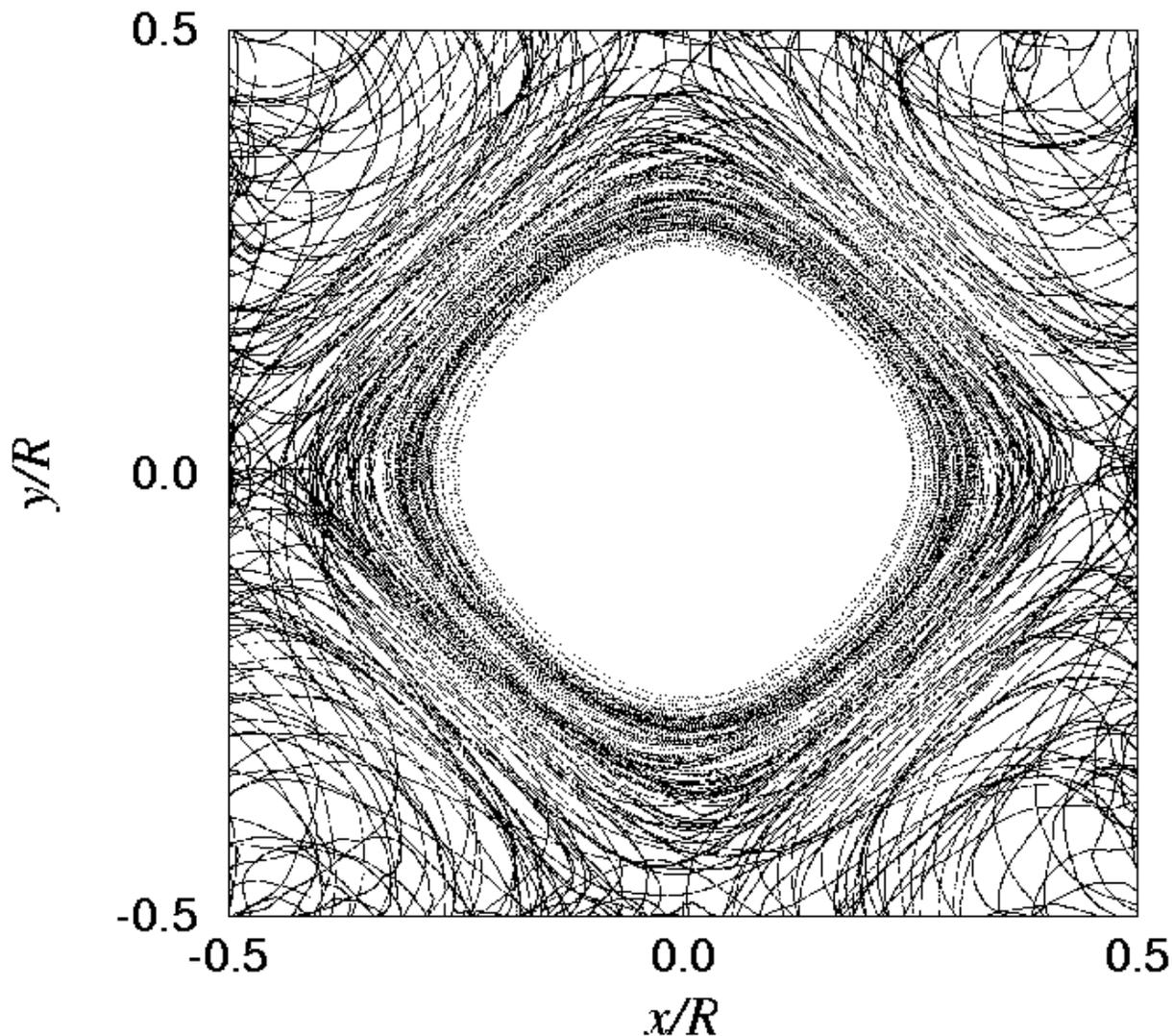}  
\caption{We show a typical long-time relative trajectory within the effective
two-body dynamics of section {\bf IV.B}. For clarity, the final time is only
$t_{\rm end}=200\,\omega^{-1}$, although we have checked that the vortices
never reach a distance less than $d_{\rm min}\approx 0.24R$, up to  $t_{\rm
end}\sim 10^5\omega^{-1}$. The anisotropy of the (square-like) trajectory is
due to the use of a Kirchhoff interaction adapted to periodic boundary
conditions. As the distance $d$ between vortices always remains of order $R$,
these effects are visible, although they should disappear when $d\ll R$ (see
discussion of section {\bf III.A}). Still, note that when vortices are close
to each other, their relative quasi-circular trajectory is not really affected
by the noise due to other vortices, as explained in section {\bf IV.C}.}
\end{figure}
\newpage 
\begin{figure}[ht]
\narrowtext
\epsfxsize=\hsize
\epsfbox{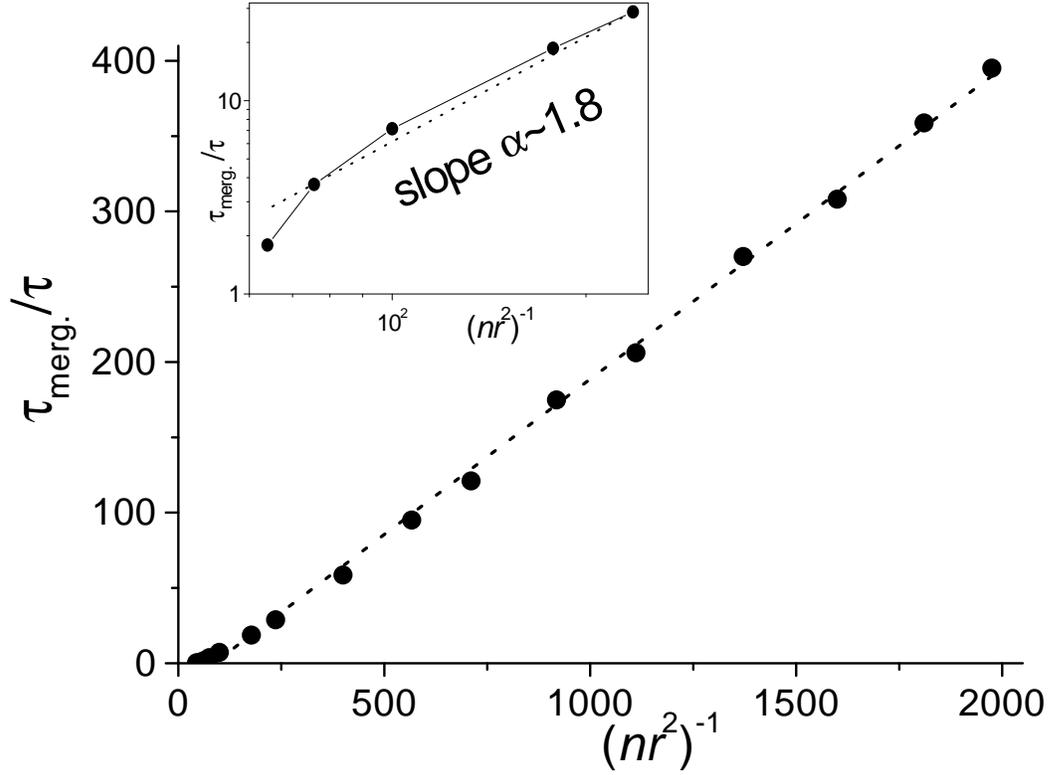}
\caption{We plot the average merging time $\tau_{\rm merg.}$ in units of the
mean free time as a function of the dimensionless parameter $(R/r)^2=
(nr^2)^{-1}$, as obtained within the effective three-body dynamics of section
{\bf IV.D}. We find a good linear behavior for small enough surface coverage.
As explained in the text, this is consistent with a decay exponent $\xi=1$.
For large surface coverage (see inset), a log-log plot leads to an apparent
exponent $\alpha\sim 1.8$, which is compatible with a decay exponent $\xi\sim
0.7$, in the language of section {\bf IV.A}.}
\end{figure}

\end{document}